\documentclass[aps,prx,reprint,showpacs,nosuperscriptaddress,floatfix]{revtex4-2}
\usepackage[utf8]{inputenc}
\usepackage{graphicx}

\usepackage{hyperref}
\hypersetup{
colorlinks=true,final=true,
        linkcolor=blue,
        citecolor=blue,
        filecolor=blue,
        urlcolor=blue,
}
\usepackage{color}
\usepackage{ulem}
\usepackage{multirow}

\begin{document}

\title{Electronic structure trends in La$_{2}R$Ni$_2$O$_7$ ($R=$ Pr, Nd, Sm) from first-principles}

\author{Yi-Feng Zhao}
 \altaffiliation {yzhao421@asu.edu}
\author{Antia S. Botana}%
\affiliation{%
Department of Physics, Arizona State University, Tempe, AZ 85287, USA}%

\date{\today}

\begin{abstract}
The discovery of superconductivity in bilayer La$_3$Ni$_2$O$_7$ under pressure has sparked tremendous attention on Ruddlesden-Popper (RP) nickelates. Recently, a higher superconducting transition temperature of 96 K was reported in Sm-doped La$_3$Ni$_2$O$_7$ single crystals at $\sim$ 22 GPa. Motivated by this experimental observation, we systematically explore the crystal structure and electronic properties of La$_3$Ni$_2$O$_7$ doped with different rare-earth elements in comparison to the undoped counterpart. As expected due to the effect of chemical pressure, we find that the volume of La$_{2}$$R$Ni$_2$O$_7$ ($R=$ Pr, Nd, Sm) progressively decreases with doping from Pr to Sm. We further find a pressure-induced structural transition to tetragonal symmetry that approximately coincides with the emergence of superconductivity in all cases. This transition is characterized by the emergence of flat $d_{z^2}$ bands at the Fermi level in the electronic structure. Despite subtle distinctions in the electronic structure between undoped and $R$-doped La$_3$Ni$_2$O$_7$, an increase in the dominant planar hopping is obtained as the $R$ size decreases. In contrast, the out-of-plane hopping decreases (in spite of the $c$ lattice constant compression), due to the decrease in the apical Ni-O$_{\rm rocksalt}$ bond length.  Our findings provide further microscopic insights into the effects of  $R$-doping in the electronic structure of RP nickelate superconductors in connection to $T_c$.

\end{abstract}
\maketitle


\section{\label{sec:level1} INTRODUCTION}

A remarkable breakthrough in the field of superconducting nickelates was marked by the observation of superconductivity with $T_c\sim$ 80 K in the bilayer Ruddlesden-Popper (RP) La$_3$Ni$_2$O$_7$ under pressures above 10-15 GPa \cite{sun2023signatures}. This compound belongs to the RP $R_{n+1}$Ni$_n$O$_{3n+1}$ ($R=$ rare-earth) nickelate family where $n$ determines the number of perovskite-like layers along the $c$-axis. Subsequently, superconductivity was discovered in other members of the RP family under similar pressure onsets including the trilayer ($n=3$) $R_4$Ni$_3$O$_{10}$ material with a $T_c\sim$ 30 K \cite{zhu2024superconductivity,zhang43102025,peipr4ni3o102026}, 
the single-layer+trilayer polymorph of La$_3$Ni$_2$O$_7$ with maximal T$_c$ up to 80 K \cite{puphal13132024,Chen2024poly,abadi13132025,huang2025superconductivity}, and the single-layer+bilayer hybrid La$_5$Ni$_3$O$_{11}$ with  $T_c\sim$ 60 K \cite{shi2025pressure}. Further, superconductivity in  thin-films of  La$_3$Ni$_2$O$_7$ upon compressive strain has also  been recently realized \cite{ko2025signatures,bhatt2025resolving,osada2025strain,li2025angle,zhou2025ambient}, albeit with lower T$_c$ than their pressurized bulk counterparts. 

Following this series of discoveries in RP nickelate superconductors, a key fundamental question to address is whether their $T_c$ can be further increased. Rare-earth doping La$_3$Ni$_2$O$_7$ has been proven to be an effective means to experimentally achieve this goal. For example, superconductivity with T$_c$ $\sim$ 82.5 K has been observed in La$_{2}$PrNi$_2$O$_7$ at P $\sim$ 18 GPa \cite{wang2024bulk}. In La$_{0.9}$Nd$_{2.1}$Ni$_2$O$_7$, a T$_c$ of up to $\sim93$ K has been achieved  at P $\sim$ 20 GPa \cite{qiu2025interlayer}.
More recently, the successful synthesis of Sm-doped La$_3$Ni$_2$O$_7$ further pushed the $T_c$ up to 96 K at P $\sim$ 22 GPa \cite{li2025bulk}. 
On the theory front, the effects of rare-earth substitution remain controversial. While some work predicted that the $T_c$ could be increased when going from La to Sm$_3$Ni$_2$O$_7$ based on the slave-boson method \cite{panrare2024}, or upon Nd-doping in La$_3$Ni$_2$O$_7$ using renormalized mean-field theory \cite{nd_doped}, other work reported a plausible decrease from La to Lu$_3$Ni$_2$O$_7$ using random phase
approximation (RPA) calculations \cite{zhangrare2023}.

In this work, we use first-principles calculations to systematically study the structural and electronic properties of the rare-earth doped bilayer RP nickelate La$_3$Ni$_2$O$_7$ and contrast them to those of the undoped compound. By evaluating the enthalpy differences for $R$ substitution on distinct La sites, we conclude that the doped atoms prefer to occupy the rocksalt layer position. We find that, across all studied materials, a pressure-induced structural transition from monoclinic to tetragonal takes place concomitantly with the onset of superconductivity.
Importantly, extra hole-like Fermi surface pockets of $d_{z^2}$ orbital character emerge at the corner of the Brillouin zone under pressure. This feature, observed in all of the compounds, has been regarded as a key ingredient for superconductivity from the perspective of electronic structure calculations. As the $R$ size decreases, the dominant in-plane hopping increases while the leading out-of-plane hopping decreases instead, due to the compressed apical Ni-O$_{\rm rocksalt}$ bond length. These trends highlight the importance of hybridization effects between the out-of-plane $d_{z^2}$ orbitals and the itinerant planar $d_{x^2-y^2}$ states in superconducting RP nickelates.

\section{\label{sec:level2}     Structural and computational details}

\begin{figure*}
    \centering
    \includegraphics[width=1.0\linewidth]{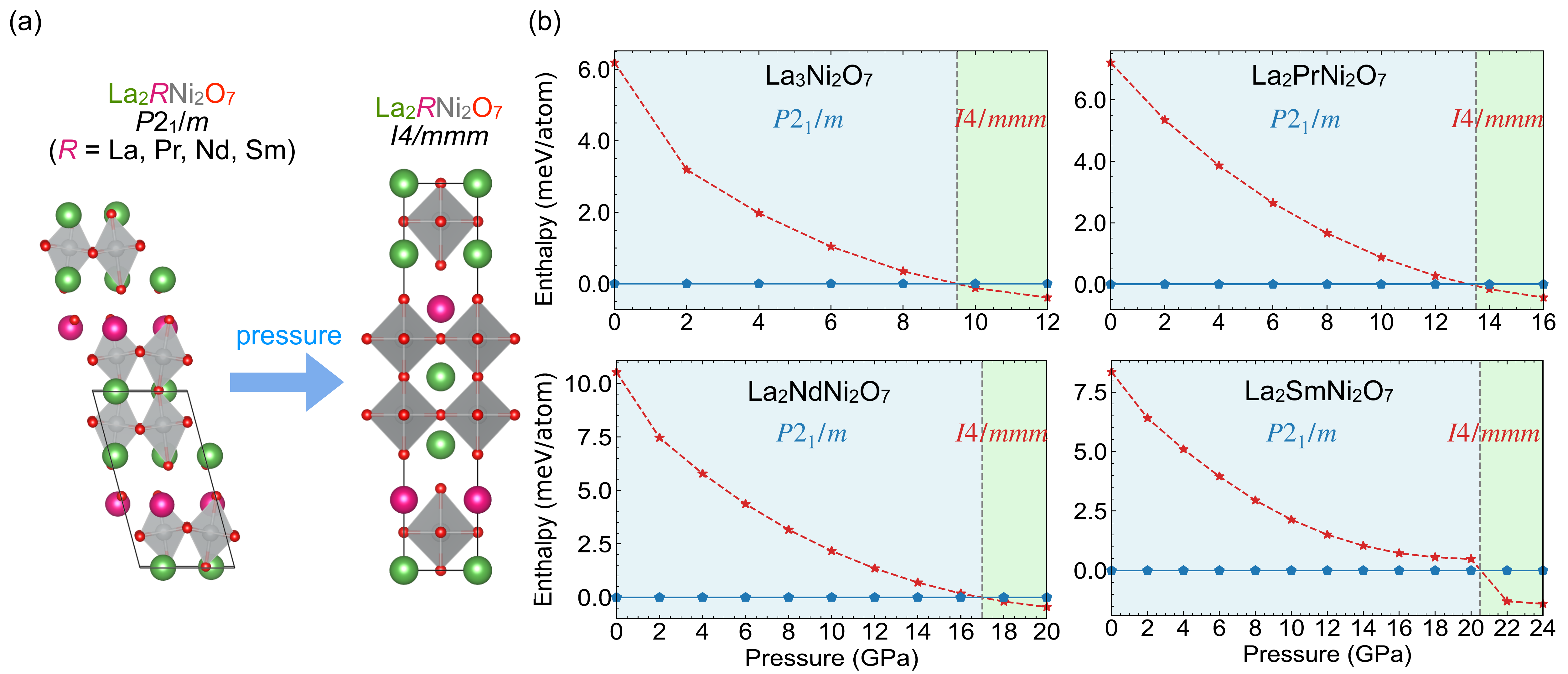}
    \caption{``Tetragonalization'' of the crystal structures of La$_2$$R$Ni$_2$O$_7$ ($R=$ La, Pr, Nd, and Sm) under pressure. (a) The ambient pressure
structure with $P2_1/m$ space group symmetry transforms to a tetragonal ($I4/mmm$) structure under pressure. The green, magenta, red, and gray spheres represent the lanthanum, $R$-doped ions, oxygen, and nickel atoms, respectively. (b)  Enthalpy difference between $P2_1/m$ and $I4/mmm$ structures for La$_2$$R$Ni$_2$O$_7$ ($R=$ La, Pr, Nd, and Sm) as a function of pressure.}
    \label{fig1}
\end{figure*}

From X-ray diffraction (XRD) data, an orthorhombic $Amam$ space group has been consistently identified  for bilayer La$_3$Ni$_2$O$_7$ at ambient pressure \cite{sun2023signatures,hou2023emergence,Yang2024arpes,dong2024visualization,zhang2024high,ZHANG2024147,wangpressure327,fan3272024,chensdw2024,wang2024structure,li327structure,shi2025spin,zhang2025identifying}. Under pressure, while initially a structural transition to $Fmmm$ symmetry was reported at 10--15\ GPa, low-temperature ($\sim$ 40 K) XRD and DFT calculations revealed an orthorhombic ($Amam$) to tetragonal ($I4/mmm$) transition instead  \cite{labollita2024electronic,wang2024structure,li327structure,geisler2024structural,zhang2025identifying}. The available XRD data for La$_2$PrNi$_2$O$_7$ show a similar pressure-induced structural transition from orthorhombic ($Amam$) to tetragonal ($I4/mmm$) symmetry takes place at $\sim$ 11 GPa, coinciding with the emergence of superconductivity \cite{wang2024bulk}. As for Nd-doped La$_3$Ni$_2$O$_7$, the same structural phase transition has been reported at a pressure $\sim$ 13 GPa for La$_{2.1}$Nd$_{0.9}$Ni$_2$O$_7$ that increases to 26 GPa for La$_{0.9}$Nd$_{2.1}$Ni$_2$O$_7$ \cite{qiu2025interlayer, fengNddoped2024}.  In contrast, the recent experimental studies on  La$_2$SmNi$_2$O$_7$ find a monoclinic $P2_1/m$ space group at ambient pressure \cite{li2025bulk} not only in the doped compound, but also in the pure La phase. Under pressure, the same type of structural transition to tetragonal $I4/mmm$ symmetry is found but at a higher pressure onset of $\sim$ 18 GPa \cite{li2025bulk}. Given that Sm-doped La$_2$SmNi$_2$O$_7$ is a prime target in our analysis and to be consistent in showing trends across all $R$ dopants, we uniformly adopt the $P2_1/m$ phase as the benchmark starting structure for all La$_2$$R$Ni$_2$O$_7$ compounds ($R=$ La, Pr, Nd, and Sm).

With these considerations in mind, we performed density-functional theory (DFT) calculations to investigate the structural properties and the electronic structure of rare-earth-doped La$_3$Ni$_2$O$_7$  employing the QUANTUM-ESPRESSO package \cite{Giannozzi_2009}. We used the Perdew-Burke-Ernzerhof (PBE) version of the generalized gradient
approximation (GGA) as the exchange-correlation functional \cite{perdew1996generalized}. To account for the strong electronic correlations in the Ni-$3d$ states, we employed the GGA$+U$ method \cite{dudarev1998electron} with an effective on-site Coulomb repulsion $U=3.5$ eV, which has been adopted in previous reports of RP nickelate superconductors to accurately describe their structural and electronic features \cite{Yang2024arpes,Shirare2025,Shaothinfile2025}. A plane-wave kinetic energy of 80 Ry and a charge density cutoff of 320 Ry were used for all the calculations.  For the Brillouin zone sampling, an $8\times8\times4$ and $12\times12\times2$ Monkhorst-Pack $k$-point grids were used for the monoclinic $P2_1/m$ and tetragonal $I4/mmm$ phases, respectively, which were enhanced to $30\times30\times2$ to refine the characteristics of the Fermi surface topology. The orbital-resolved Fermi surfaces were visualized using the FermiSurfer code \cite{kawamura2019fermisurfer}.

The initial structural data of undoped La$_3$Ni$_2$O$_7$ in the $P2_1/m$ and $I4/mmm$ space groups was taken from Refs.~\cite{li2025bulk} and \cite{wang2024structure}, respectively. Single crystal X-ray diffraction measurements indicate that Sm and Pr dopants prefer to occupy the La sites in the rocksalt layer \cite{wang2024bulk,li2025bulk}, as illustrated in Fig.~\ref{fig1}(a). We independently verified that this substitution is the most energetically favorable one. Hence, in the doped systems, we replaced two La atoms in the rocksalt layers by  $R=$~Pr, Nd, and Sm in the conventional unit cell (that contains two bilayer units), to form the La$_2$$R$Ni$_2$O$_7$ structures. We then fully optimized all the crystal structures explored (both unit cell and internal coordinates) until the residual forces were below $10^{-4}$ Ry/Bohr. The threshold of scf-consistent calculations is set to $10^{-10}$ Ry.

In order to analyze trends in relevant hoppings, we constructed maximally localized Wannier functions (MLWFs) by employing Wannier90 \cite{Pizzi2020}. The Ni-$e_{g}$ orbitals ($d_{x^2-y^2}$ and $d_{z^2}$) were chosen as the projection basis to downfold the DFT results onto MLWFs.

\section{\label{sec:level3} RESULTS AND DISCUSSION}

\subsection{Structures}

Figure~\ref{fig1}(a) displays the crystal structure of La$_2R$Ni$_2$O$_7$ in $P2_1/m$ symmetry, which clearly shows the series of two NiO$_6$ perovskite-like planes separated by rocksalt layers along the $c$ axis. To determine a stable doped structure at ambient pressure, we compare the enthalpy difference between two distinct La$_2R$Ni$_2$O$_7$ configurations, where the $R$ atoms occupy the outer (rocksalt) layers  or the inner layers (perovskite blocks). Our results (see Appendix~\ref{appendix:A}) confirm that rocksalt substitution is energetically preferred, in agreement with experiments and previous DFT work \cite{wang2024bulk,li2025bulk, huoprdoed2025}. Consequently, we adopt this configuration (as shown in Fig.~\ref{fig1}(a)) as the starting structure type to perform systematic comparisons in La$_2$$R$Ni$_2$O$_7$ ($R=$ La, Pr, Nd, and Sm).

Subsequently, we study the possibility of a structural transition with pressure. The evolution with pressure of the enthalpy difference between monoclinic $P2_1/m$ and tetragonal $I4/mmm$ phases is shown in Fig.~\ref{fig1}(b). The calculated transition pressures from $P2_1/m$ to $I4/mmm$  for undoped La$_3$Ni$_2$O$_7$ and for La$_2$$R$Ni$_2$O$_7$ with $R=$~Pr, Nd, and Sm, are 9.5 GPa, 14 GPa, 17 GPa, and 20.5 GPa, respectively. These critical pressures are qualitatively in good agreement with experiments where the transition occurs at 11 GPa for Pr doping \cite{wang2024bulk}, 13 GPa for Nd doping \cite{qiu2025interlayer}, and 18 GPa for Sm doping \cite{li2025bulk}. 
The gradual increase in transition pressure arises from the decreasing ionic radius of the doped rare-earth elements, and the concomitant effect of chemical pressure.  

\begin{figure}
    \centering
    \includegraphics[width=1.0\linewidth]{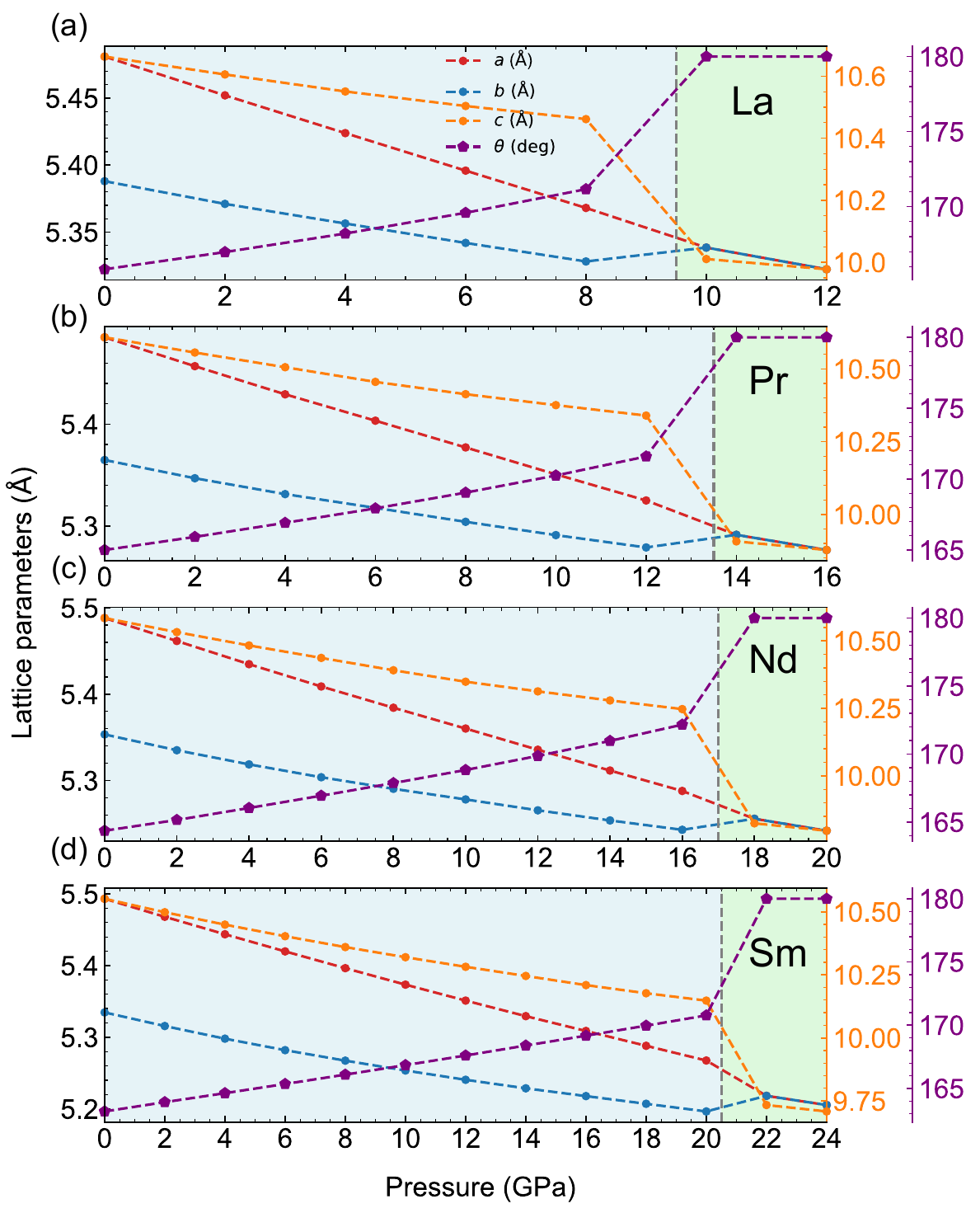}
    \caption{Structural data (lattice constants and apical Ni-O-Ni bond angles) capturing the structural transition to tetragonal symmetry as a function of pressure in La$_{2}$$R$Ni$_2$O$_7$ ($R$ = La (a), Pr (b), Nd (c), Sm (d)); with $a$ becoming equal to $b$, and the apical bond angle turning 180$^\circ$.}
    \label{fig2}
\end{figure}

To gain a deeper understanding into the changes that take place in this structural transition, we show the evolution of the lattice parameters as well as of the apical Ni-O-Ni bond angle for undoped and doped La$_3$Ni$_2$O$_7$ as a function of pressure in Fig. \ref{fig2}. 
From undoped La$_3$Ni$_2$O$_7$ (panel (a)) to La$_2$SmNi$_2$O$_7$ (panel (d)), at ambient pressure, the lattice constant $a$ increases slightly from 5.481 \AA~ for the undoped bilayer material to 5.493 \AA~ for the Sm-doped compound, while $b$ and $c$ decrease from 5.388 to 5.335 \AA~and from 10.663 to 10.552 \AA, respectively. These tendencies are in line with reports on  La$_2$PrNi$_2$O$_7$ \cite{wang2024bulk,huoprdoed2025}, La$_{3-x}$Nd$_x$Ni$_2$O$_7$ (0.3 $\leq x \leq$ 2.4) \cite{qiu2025interlayer}, as well as La$_2$SmNi$_2$O$_7$ \cite{li2025bulk}. As for the apical Ni-O-Ni bond angle, it decreases from 165.8$^\circ$ for undoped to 163.2$^\circ$ for Sm-doped La$_3$Ni$_2$O$_7$, also in agreement with experimental trends \cite{li2025bulk}. 

In the high-pressure tetragonal phase, the tendencies are similar to those at ambient pressure in terms of volume changes. At 21 GPa, all planar lattice parameters $a$ ($b$) and the out-of-plane parameter $c$ decrease from 3.717 to 3.695 \AA~and from 19.691 to 19.493 \AA,  respectively, when going from pure La to the Sm-doped compound. We pick 21 GPa as all compounds are tetragonal at this pressure, regardless of the rare-earth dopant, with apical Ni-O-Ni bond angles of 180$^\circ$, similar to the effect introduced by hydrostatic pressure.  

\subsection{Electronic Structure}

\begin{figure}
    \centering    \includegraphics[width=1.0\linewidth]{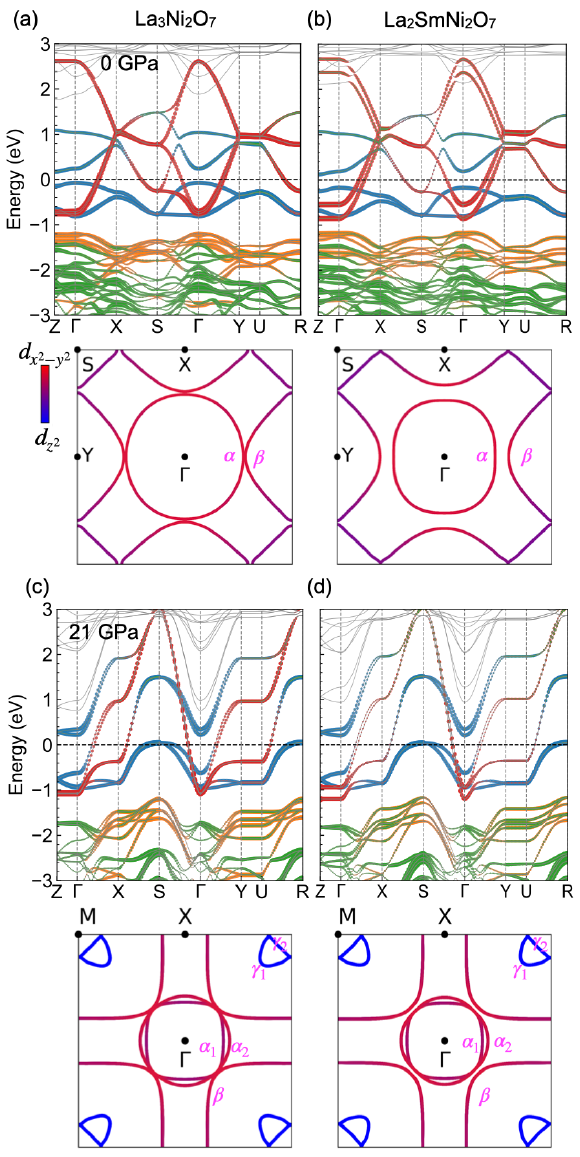}
    \caption{Electronic band structures  and corresponding Fermi surfaces in the $k_{z}=0$ plane for La$_2$$R$Ni$_2$O$_7$ ($R=$ La and Sm) at ambient pressure (a,b) and at 21 GPa (c,d).}
    \label{fig3}
\end{figure}

We continue our analysis by looking at the changes in the nonmagnetic electronic structure induced by $R$ doping in La$_3$Ni$_2$O$_7$ at ambient pressure. The electronic structure of La$_3$Ni$_2$O$_7$ has been widely investigated at both ambient and high pressure in previous work \cite{luo2023bilayer,Zhang2023,Yang2023,lechermann2023electronic,christiansson2023correlated,zhang2024structural,labollita2024electronic,ouyang2024absence,hanghui2024electronic}. 
The nominal Ni valence for La$_3$Ni$_2$O$_7$ is 2.5+, corresponding to an average $d^{7.5}$ filling for the Ni ions. With the Ni $t_{2g}$ orbitals being fully occupied, there are 1.5 remaining electrons to be accommodated in the $e_g$ orbitals per Ni. As a consequence (as shown in Fig.~\ref{fig3}(a)) at ambient pressure the Ni-$d_{z^2}$ and $d_{x^2-y^2}$ states are both active near the Fermi level. The $d_{z^2}$ bands are split into a bonding and antibonding molecular-orbital combination (by $\sim$ 1 eV at the $\Gamma$ point) due to the quantum confinement of the NiO$_6$ bilayer block in the structure.  The Ni-$d_{x^2-y^2}$ states have a large bandwidth $\sim$ 3.4 eV. The $t_{2g}$ states and the $e_g$ states are well separated by $\sim$ 0.3 eV. 

The impact of $R$-doping on the electronic structure near the Fermi level is subtle but non-negligible (we show in Fig.~\ref{fig3}(b) the Sm doped case, the other $R$s are shown in Appendix \ref{appendix:B}). The main changes are: (i) The $d_{z^2}$ bonding band near the Fermi level at the $\Gamma$ point is pushed down in energy. This effect is similar to that obtained under compressive strain \cite{zhao2025electronic,huo2025modulation,geisler2025}. (ii) The bandwidth of the $d_{x^2-y^2}$ bands increases by $\sim$ 0.3 eV for La$_2$SmNi$_2$O$_7$, as expected. Further, an energy splitting in the Ni-$d_{x^2-y^2}$ bands arises and increases monotonically as the $R$ size decreases, as a consequence of $R$ substitution that lowers the local symmetry. 

The electronic band structures at 21 GPa (Fig.~\ref{fig3}(c), (d)) are remarkably distinct from those at ambient pressure. The splitting between the Ni-$d_{z^2}$ bonding and antibonding states increases by $\sim$ 0.5 eV compared to that at ambient pressure. More importantly, the flat $d_{z^2}$ bonding bands now cross the Fermi level at the $S$ point of the Brillouin zone. The metallization of the $d_{z^2}$ bonding bands and the emergence of the associated hole pockets has been considered a critical electronic signature of superconductivity at high pressure in bilayer RP nickelates \cite{sun2023signatures,lechermann2023electronic,zhang2024structural,Lu2024}. Further, the bandwidth of the Ni-$d_{x^2-y^2}$ bands increases to $\sim$ 4 eV (and is larger in the Sm case).

Figure~\ref{fig3} presents the accompanying Fermi surfaces at ambient pressure (upper panels) and at 21 GPa (lower panels).  
At ambient pressure, the Fermi surface of La$_3$Ni$_2$O$_7$ is characterized by an electron-like sheet ($\alpha$) and a hole-like sheet ($\beta$). The former pocket is predominantly Ni-$d_{x^2-y^2}$ in character, while the other has a clear admixture of Ni-$d_{z^2}$ orbital character. These features align with previous ARPES data \cite{Yang2024arpes}. In doped La$_3$Ni$_2$O$_7$, the dominant pockets remain similar to those of the undoped case. As the ionic radius of the $R$ dopant decreases, however, the sizes of $\alpha$ and $\beta$ pockets change, linked to the band shifts described above.

The Fermi surface at 21 GPa consists of five pockets instead (note that we look here at the conventional unit cell, the corresponding primitive cell results can be seen in Appendix \ref{appendix:C}). There are two electron-like pockets ($\alpha_1$ and $\alpha_2$), and three hole-like pockets ($\beta$, $\gamma_1$, and $\gamma_2$). 
The extra hole pockets ($\gamma_1$ and $\gamma_2$), which are absent at ambient pressure, are predominantly $d_{z^2}$ in character and arise from the flat $d_{z^2}$ bonding band that crosses the Fermi level at the $S$ point in the electronic band structure. 
In contrast, the $\alpha$ and $\beta$ pockets have dominant $d_{x^2-y^2}$ orbital character. While some small changes can be observed upon $R$-doping, the Fermi surface topology essentially remains.

\begin{figure}
    \centering
    \includegraphics[width=1.0\linewidth]{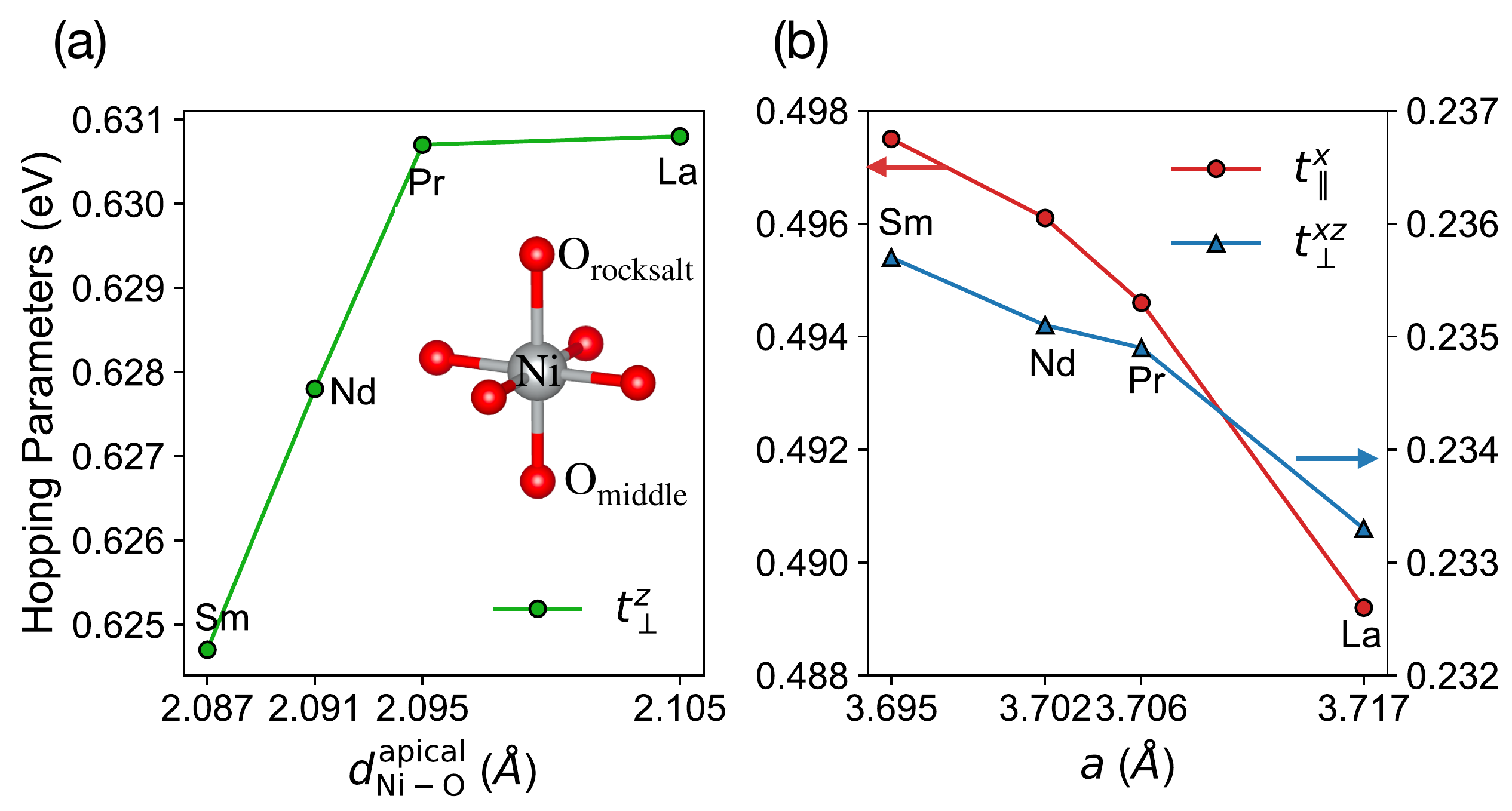}
    \caption{Relevant hopping parameters extracted from MLWFs as a function of rare-earth doping in La$_2$$R$Ni$_2$O$_7$ at high pressure. $t_{\perp}^z$ (panel (a)) represents the hopping between out-of-plane Ni-$d_{z^2}$ orbitals-- it is plotted with respect to the average Ni-O apical distance. $t_{\Vert}^x$ represents the hopping between planar Ni-$d_{x^2-y^2}$ orbitals, $t_{\Vert}^{xz}$ denotes the hopping between Ni-$d_{x^2-y^2}$ and $d_{z^2}$ orbitals. These two hoppings are plotted in panel (b) with respect to the in-plane lattice constant.}
    \label{fig6}
\end{figure}

To further understand how  $R$-doping influences the electronic structure, we also analyze the corresponding evolution of the dominant hopping parameters extracted from wannierizations. In La$_3$Ni$_2$O$_7$ upon hydrostatic pressure, there is an increase in the interlayer coupling within the bilayer   ($t^{z}_{\perp}$) arising from the large overlap between Ni-$d_{z^2}$ and O-$p_z$ orbitals \cite{zhang2024structural, lechermann2023electronic,luo2023bilayer,hanghui2024electronic,Lu2024,xie2024strong,yicdw2024,labollita_dw, jistrongcoupling3272025,yiscdw2025}. The dominant planar hopping integral ($t_{\Vert}^{x}$) also increases and is primarily related to the Ni-$d_{x^2-y^2}$ orbital overlap. In addition, the hybridization between Ni-$d_{z^2}$ and $d_{x^2-y^2}$ orbitals, represented by $t_{\Vert}^{xz}$, has been shown to be sizable. 

Figure \ref{fig6}(a,b) shows the evolution of these dominant hoppings  with chemical pressure in  La$_2$$R$Ni$_2$O$_7$. We present the values obtained at 21 GPa as a function of $R$. In contrast to hydrostatic pressure, $t^{z}_{\perp}$ decreases with chemical pressure, with the Sm-doped compound having the smallest $t^{z}_{\perp}$. We relate this reduction with the change in the apical Ni-O distance for the oxygen in the rocksalt layer. Upon $R$ substitution, while the Ni-O apical distance within the bilayer block (Ni-O$_{\rm middle}$) remains practically unchanged, the apical Ni-O$_{\rm rocksalt}$ bond length decreases significantly (see Appendix \ref{appendix:E}).  Hence, the compression in $c$ obtained for smaller $R$ is mostly absorbed by the rocksalt slab. It is the Ni-O bond length to the apical oxygen in the rocksalt that clearly controls $t^{z}_{\perp}$, that increases as this bond length increases or, in other words, as the bilayer unit is isolated along $c$. In contrast, as expected, the $t_{\Vert}^{x}$ increases as the size of the $R$ dopant decreases (as the in-plane lattice constant is compressed). The hybridization between the Ni-$d_{z^2}$ and $d_{x^2-y^2}$ orbitals also increases as the Ni-O planar distance gets smaller. Overall, the hoppings we derived:  i) caution against directly relating a reduction in $c$ upon $R$ doping with an increase in $t^{z}_{\perp}$ (and indirectly in $T_c$); ii) highlight the relevant role of the apical Ni-O$_{\rm rocksalt}$ bond length over the  dominant $t^{z}_{\perp}$ hopping; iii) emphasize the importance of hybridization effects between the off-plane $d_{z^2}$ orbitals and the itinerant planar $d_{x^2-y^2}$ states in superconducting RP nickelates.

\section{CONCLUSIONS}
In summary, we have performed a systematic investigation of the structural and electronic properties of $R$ doping in La$_3$Ni$_2$O$_7$ as a function of hydrostatic pressure via first-principles calculations. The doped atoms preferentially occupy the La sites in the LaO rocksalt slab, in agreement with experiments. All compounds undergo a pressure-driven structural transition from a monoclinic to a tetragonal phase, with the critical pressure of this transition increasing as the $R$ size decreases. The electronic structure evolution (both at ambient and under pressure) shows only subtle modifications with $R$: band shifts and corresponding changes in Fermi surface pocket size. At high pressure the most important change in the electronic structure is the emergence of flat bands of $d_{z^2}$ character at the Fermi level, in analogy to hydrostatic pressure. However, in contrast to hydrostatic pressure, the dominat $t_{\perp}^z$ hopping decreases upon chemical pressure (as $R$ decreases). This change is associated with the decrease in apical Ni-O bond length to the rocksalt layer. The dominant planar hopping  instead increases upon chemical pressure. Given that the T$_c$ of bilayer RPs been observed to increase with chemical pressure, our results seem to emphasize the importance of hybridization effects between the out-of-plane $d_{z^2}$ orbitals and the itinerant planar $d_{x^2-y^2}$ states in relation to superconductivity in this family of materials.

\section{acknowledgments}

We acknowledge support from NSF Grant No. DMR-2323971, as well as the ASU Research Computing Center for HPC resources.

\onecolumngrid
\appendix

\section{\label{appendix:A} Energy comparison for distinct $R$ substitution sites}

Table \ref{tab:energy} shows the energy difference for $R$-doping in the inner (perovskite) layer vs. the outer (rocksalt) slab. $R$-substitution in the latter is energetically favorable. 

\begin{table}[h]
    \centering
        \caption{Relative energy (in eV) for Sm-doping in the perovskite layer and in the rocksalt slab for La$_2$SmNi$_2$O$_7$.}
    \setlength{\tabcolsep}{22pt}
    \begin{tabular}{ccc}
    \hline
    \hline
        Substitution-site & Inner (perovskite) & Outer (rocksalt) \\
    \hline
        Energy & 0.554 & 0 \\
    \hline
    \hline
    \end{tabular}
    \label{tab:energy}
\end{table}

\section{\label{appendix:B} Electronic structure for La$_2$$R$Ni$_2$O$_7$ ($R=$ Pr, and Nd)}

\begin{figure*}
    \centering
    \includegraphics[width=0.7\linewidth]{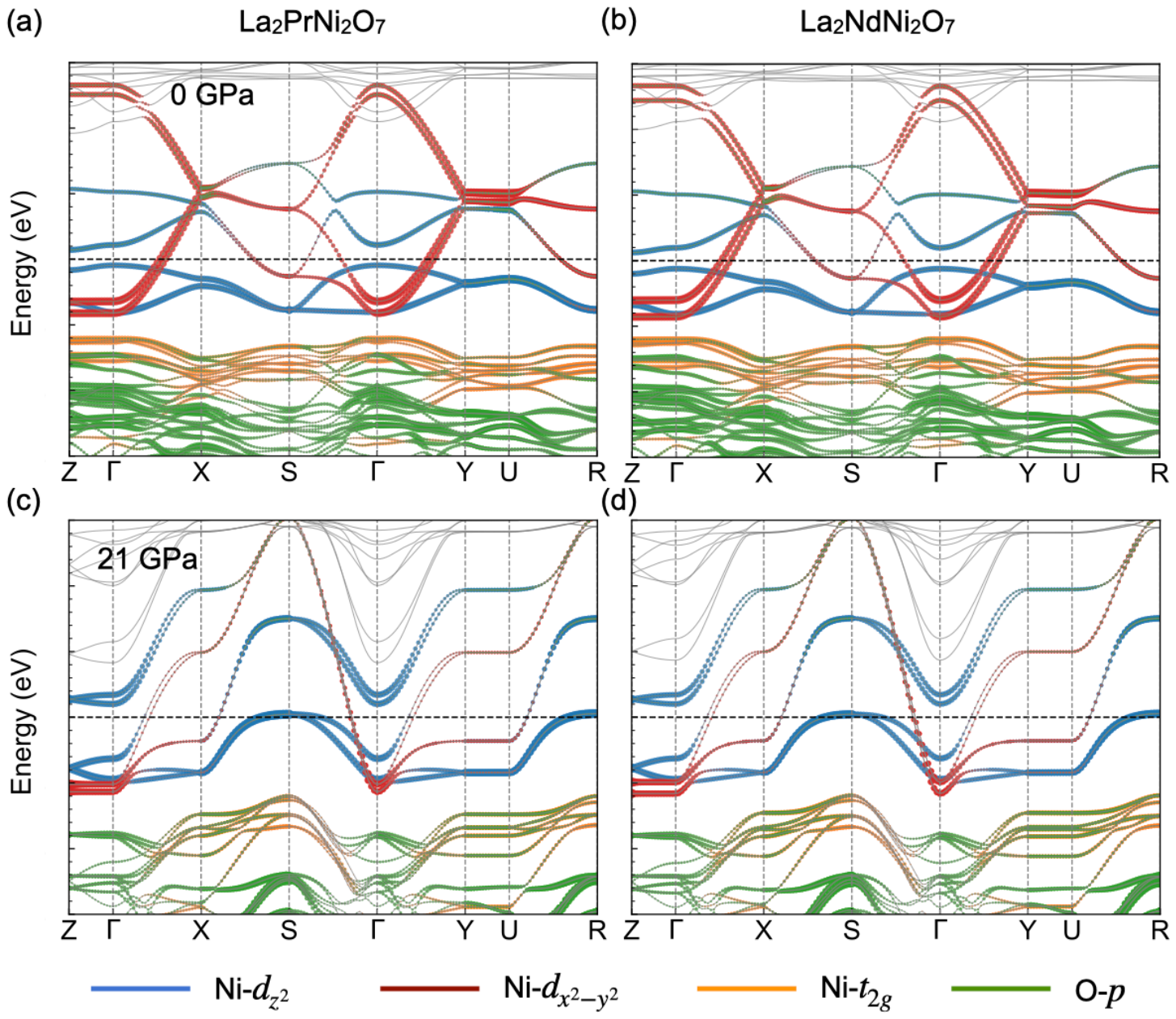}
    \caption{Electronic band structures for La$_2$$R$Ni$_2$O$_7$ ($R=$  Pr, Nd) at ambient pressure and high pressure with orbital character highlighted.}
    \label{figs1}
\end{figure*}

The band structures for La$_2$PrNi$_2$O$_7$ and La$_2$NdNi$_2$O$_7$ at ambient pressure and 21 GPa are shown in Fig. \ref{figs1}. The same trends as those described in the main text for La$_3$Ni$_2$O$_7$ and La$_2$SmNi$_2$O$_7$ are observed.

\begin{figure*}
    \centering
    \includegraphics[width=0.5\linewidth]{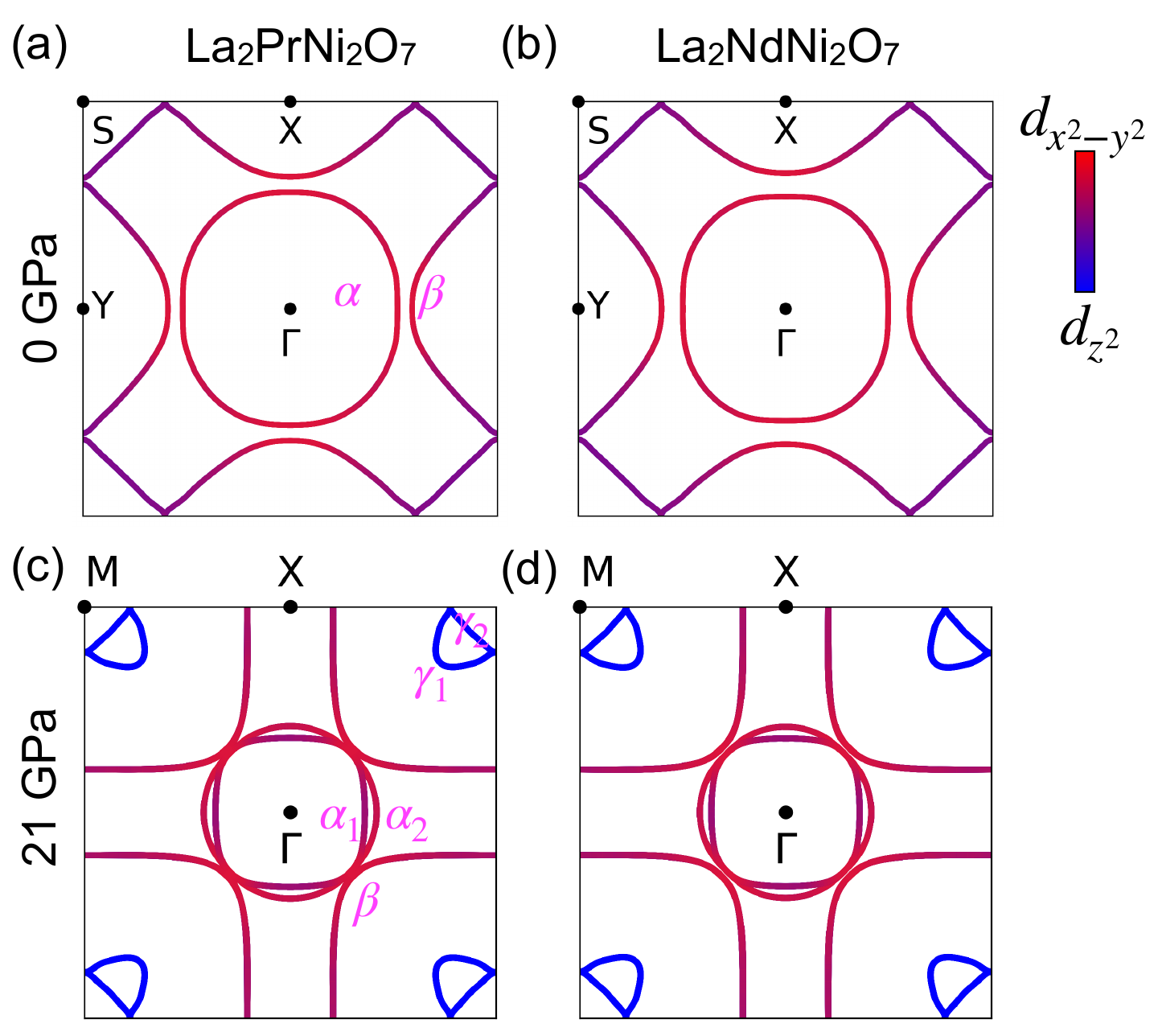}
    \caption{Fermi surfaces for La$_2$$R$Ni$_2$O$_7$ ($R=$ Pr, Nd) at ambient pressure and high pressure with orbital character highlighted.}
    \label{figs2}
\end{figure*}

Figure \ref{figs2} shows the corresponding Fermi surfaces for La$_2$PrNi$_2$O$_7$ and La$_2$NdNi$_2$O$_7$ at ambient pressure and 21 GPa. The same tendencies as those described in the main text are once again retained. The emergence of extra $\gamma$ pockets at 21 GPa is also observed. 

\begin{figure}
    \centering
    \includegraphics[width=0.5\linewidth]{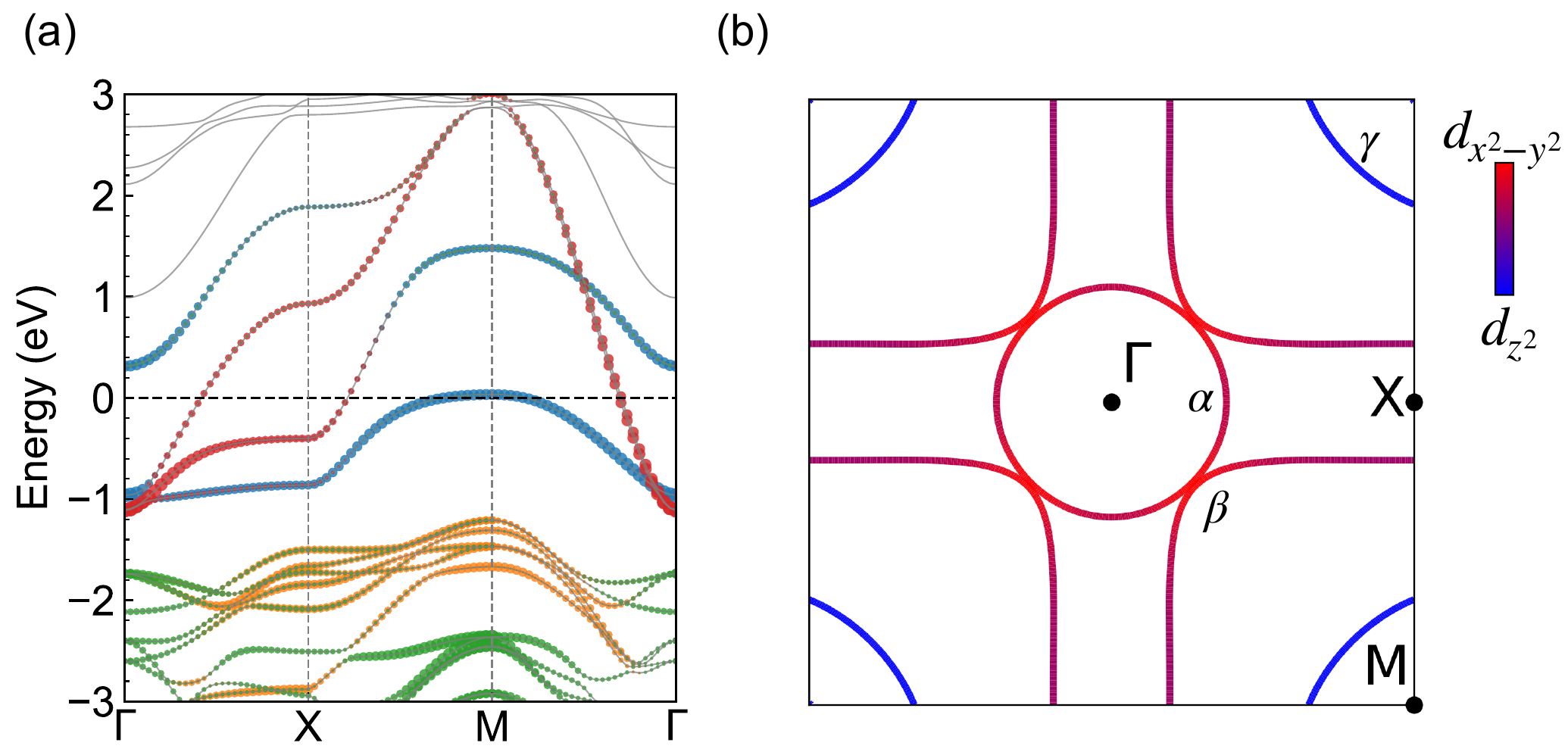}
    \caption{Electronic band structure and Fermi surface of La$_3$Ni$_2$O$_7$ in the primitive unit cell at 21 GPa. The red and blue colors represent Ni-$d_{x^2-y^2}$ and $d_{z^2}$ orbitals.}
    \label{figs4}
\end{figure}

\section{\label{appendix:C} Electronic band structure for La$_3$Ni$_2$O$_7$ in the primitive cell under pressure}

Figure \ref{figs4} shows the electronic band structure and Fermi surface of La$_3$Ni$_2$O$_7$ in  the primitive unit cell at 21 GPa. The Fermi surface shows a single $\alpha$, $\beta$ and $\gamma$ pockets. 

\section{\label{appendix:E} Further structural data for rare-earth doped La$_3$Ni$_2$O$_7$}

\begin{figure}
    \centering
    \includegraphics[width=0.8\linewidth]{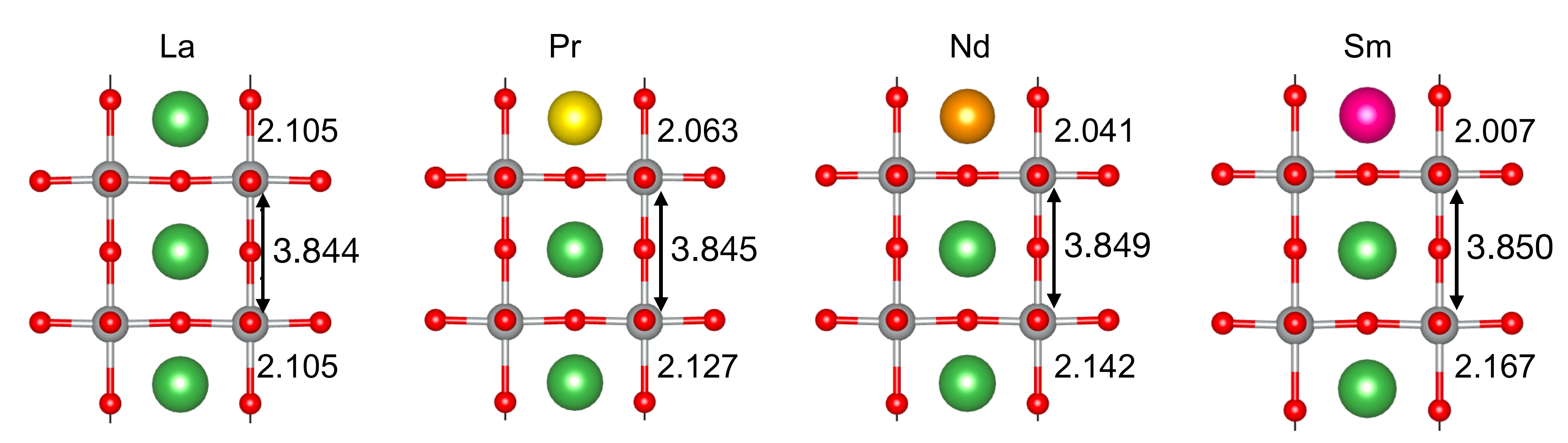}
    \caption{Out-of-plane bond length changes in rare-earth doped La$_3$Ni$_2$O$_7$ at 21 GPa.}
    \label{figs5}
\end{figure}

Figure \ref{figs5} shows changes in relevant bond lengths in La$_2$$R$Ni$_2$O$_7$ ($R$= La, Pr, Nd, Sm). The average out-of-plane Ni-O distance within the bilayer remains practically unchanged with $R$, while the out-of-plane Ni-O distance to the rocksalt decreases significantly with the $R$ size. 

\clearpage
\twocolumngrid
\bibliography{main}

\end{document}